\documentclass[sigconf,authorversion]{acmart}

\usepackage{hyperref}
\usepackage{url}
\usepackage{tikz}
\usepackage{float}
\usepackage{xcolor}

\usepackage{balance}

\usetikzlibrary{arrows.meta,positioning,shapes.geometric,fit}

\newcommand{\revision}[1]{\textcolor{black}{#1}}
\newcommand{\revisiontwo}[1]{\textcolor{black}{#1}}
\newcommand{\revisionthree}[1]{\textcolor{black}{#1}}
\newcommand{\revisioncrc}[1]{\textcolor{black}{#1}}
\newcommand{\AAF}{Agency Allocation Framework (AAF)}
\newcommand{\AAFshort}{AAF}

\AtBeginDocument{%
}

\copyrightyear{2026}
\acmYear{2026}
\setcopyright{cc}
\setcctype{by}
\acmConference[L@S '26]{Proceedings of the Thirteenth ACM Conference on Learning @ Scale}{June 29-July 03, 2026}{Seoul, Republic of Korea}
\acmBooktitle{Proceedings of the Thirteenth ACM Conference on Learning @ Scale (L@S '26), June 29-July 03, 2026, Seoul, Republic of Korea}
\acmDOI{10.1145/3774398.3811627}
\acmISBN{979-8-4007-2293-6/2026/06}

\begin{document}

\title[The Agency Allocation Framework]{Who Decides in AI-Mediated Learning?}
\subtitle{The Agency Allocation Framework}

\author{Conrad Borchers}
\orcid{0000-0003-3437-8979}
\affiliation{
    \institution{Carnegie Mellon University}
    \city{Pittsburgh}
    \state{PA}
    \country{USA}
}
\email{cborcher@cs.cmu.edu}

\author{Olga Viberg}
\orcid{0000-0002-8543-3774}
\affiliation{
    \institution{KTH Royal Institute of Technology}
    \city{Stockholm}
    \country{Sweden}
}
\email{oviberg@kth.se}

\author{René F. Kizilcec}
\orcid{0000-0001-6283-5546}
\affiliation{
    \institution{Cornell University}
    \city{Ithaca}
    \state{NY}
    \country{USA}
}
\email{kizilcec@cornell.edu}

\renewcommand{\shortauthors}{Conrad Borchers, Olga Viberg, and René F. Kizilcec}

\begin{abstract}
As AI-mediated learning systems increasingly shape how learners plan, make decisions, and progress through education, learner agency is becoming both more consequential and harder to conceptualize at scale. Existing research often treats agency as a proxy for engagement and self-regulation, leaving unclear who actually holds decision-making authority in large-scale, automated learning environments. This paper reframes learner agency as the allocation of decision authority across learners, educators, institutions, and AI systems. We introduce the Agency Allocation Framework (AAF) for analyzing how decisions are distributed, how choices are architected, what evidence supports them, and over what time horizons their consequences unfold. Drawing on a focused review of Learning at Scale literature and an illustrative tutoring-system example, we identify four recurring challenges for studying learner agency at scale: (1) conceptual ambiguity, (2) reliance on behavioral proxies, (3) trade-offs between efficiency and learner control, and (4) the redistribution of agency through AI-mediated systems. Rather than advocating more or less automation, the AAF supports systematic analysis of when AI scaffolds learners' capacity to act and when it substitutes for it. By making decision authority explicit, the framework provides researchers and designers with analytic tools for studying, comparing, and evaluating agency-preserving learning systems in increasingly automated educational contexts.
\end{abstract}



\maketitle

\section{Introduction}

\revision{
Student agency is commonly defined as learners' capacity to act intentionally and make choices aligned with their goals \cite{brod2026agency}. In educational research, this capacity is rarely understood as purely individual. Instead, definitions of learner agency emphasize the social, institutional, and material conditions that enable or constrain learners' ability to act (e.g., \cite{mairitsch2023putting, yung2025theorizing}). From this perspective, agency is conceptualized as a dynamic and relational capacity through which learners initiate, regulate, and reflect on their learning by setting and pursuing valued goals, exercising self-regulation, and adapting action in response to contextual constraints \cite{OECD2018LearnerAgency}. These accounts are grounded in broader theories of human agency, which characterize agency through interrelated processes: intention, forethought, self-regulation, and self-reflection \cite{bandura2006toward}. Across these traditions, agency is consistently enacted within enabling and constraining environments. Yet, they differ in where agency is assumed to reside, how it is enacted, and, critically, what counts as evidence of agency.
}

\revision{
These differences become especially consequential in learning environments that operate at scale. Large-scale systems rely on automation, standardization, and algorithmic mediation to function, thereby fundamentally changing the forms of interaction available to learners \cite{roll2018learning}. As a result, control and decision-making regarding planning, sequencing, feedback, and pathway navigation are often partially relinquished by learners and redistributed to platforms, policies, and AI-driven systems \cite{kizilcec2023pipelines}. Under these conditions, learner agency can be easily overlooked once control has been relinquished. Not because it is absent, but because it is exercised through system-structured decisions rather than directly observable acts of choice.}

\revision{
Much of the existing empirical work on learner agency has been conducted in small-scale, human-mediated settings or has inferred agency post hoc from engagement-related signals. Prior review papers, for example, examine taxonomies of agentic processes in learners' engagement with feedback \cite{winstone2017supporting} or draw on empirical studies of small-group learning processes \cite{tay2025role}. These assumptions do not readily extend to learning at scale, where agency must be understood in relation to the options systems make available, the constraints they impose, and the decisions they structure across extended learning trajectories. As we show in Section~\ref{sec:lit-review}, our review of Learning at Scale conference proceedings and agency-related terminology finds that scholars have not systematically reviewed or advanced a consistent framework for agency. The term has been used inconsistently to refer to distinct underlying theoretical constructs (e.g., learner control, self-directed learning, help-seeking). This presents both a challenge and an opportunity: a challenge because agency is harder to observe under conditions of automation and limited researcher orchestration, and an opportunity because system-level design decisions render explicit (and consequential) how learners' opportunities to act are shaped and unevenly distributed over time. We respond by proposing a novel perspective on the platform-learner relationship.
} \revisiontwo{We discuss how this perspective can enhance the research and design of technology-mediated or distributed learning. The \AAFshort{} summarizes practical guidance and implications for learning at scale and adjacent communities (e.g., AIED, learning analytics) discussed later.}

\subsection{Agency: Interplay of Platform and Learner}

\revision{As AI-mediated systems increasingly participate in learners' planning, forethought, and decision-making, the empirical conditions under which learner agency can be studied are changing. Only a few studies have investigated how researchers can measure and foster students' literacy and self-efficacy when using generative AI during learning \cite{anders2025developing}. More capable AI systems actively shape choices, priorities, and pathways, occupying decision-making roles that humans previously held \cite{floridi2024ai}. Under these conditions, questions of learner agency (i.e., who plans, who takes action, and on what basis) become more consequential than in earlier generations of educational technology. Further, large-scale learning environments generate unprecedented volumes of behavioral data, spanning micro-level task interactions to macro-level academic pathways such as degree progression \cite{fischer2020mining}. These data can serve as a rich empirical basis for studying and fostering learner agency, making visible patterns of choice, persistence, regulation, and adaptation at fine granularity, for instance, through video and audio recordings of tutoring~\cite{borchers2026brief,kizilcec2023pipelines}.}

\revision{
Taken together, these developments highlight the need for a clearer way to reason about learner agency in AI-mediated learning environments. While behavioral and multimodal data make aspects of learners' regulation increasingly visible \cite{jarvela2021multimodal}, they do not, on their own, resolve how agency should be conceptualized, located, or evaluated when decisions are distributed across learners and systems. In this position paper, we adopt the lens that learner agency at scale is enacted through an interplay between learners and learning infrastructures in a given learning setting. We argue that this interplay must be analyzed at the level of concrete decisions: what decisions are made, by whom, under what constraints, and over what time horizons. To support this shift, we introduce the \AAF{} to provide a structured approach for studying, designing for, and evaluating learner agency under conditions of automation, partial observability, and scale.}

\revision{
This framing distinguishes our contribution from prior work on learner agency, which has largely focused on individual dispositions or small-scale instructional settings \cite{brod2026agency, anders2025developing}. Rather than proposing a new definition or measure of agency, we introduce the \AAFshort{} for analyzing how agency is produced, constrained, and redistributed in large-scale learning systems. The \AAFshort{} provides researchers with a common language for designing studies and comparing systems based on how decision-making authority is structured across learners and learning infrastructures.}

\subsection{Timeliness}

\revision{
As learning systems scale, learners may cede planning, monitoring, and decision-making to automated and AI-mediated supports that increasingly shape what learning options are available and salient \cite{floridi2024ai}. Emerging evidence suggests a potential downside: when learners rely heavily on highly capable systems, they may offload regulatory activities, thereby reducing opportunities to develop self-regulation and related learning capacities \cite{fan2025beware}.
\revisionthree{Evidence is still accumulating, especially for at-scale settings, but the risk motivates targeted empirical tests.}
Because any attenuation of agency may unfold gradually and unevenly, large-scale environments can appear successful on conventional metrics (e.g., engagement) while narrowing learners' capacity to plan, choose, and adapt once system support is reduced. This makes learner agency a timely and underexamined concern for Learning at Scale research and motivates the need for analytic tools that make decision authority explicit.}

As the field develops tools that influence longer-term academic trajectories (e.g., course selection) \cite{kizilcec2023pipelines, malik2025transcripts, thompson2024gender}, assumptions about learner agency embedded in system logic become increasingly consequential. This paper responds by introducing the \AAFshort{} and a set of concrete challenges to help scholars examine how agency is shaped, constrained, and redistributed in large-scale systems. Learner agency is also becoming a regulatory concern: frameworks such as the EU's General Data Protection Regulation (GDPR) \cite{gdpr2018general} and the AI Act \cite{act2024artificial}, emphasize autonomy, transparency, and protections around automated decision-making in high-stakes domains, including education. Hence, operationalizing learner agency also becomes a compliance and human rights design imperative \cite{hutt2023right}. The \AAFshort{} provides a structured way to analyze these allocations of decision-making authority and their consequences over time.

\subsection{The Present Study and Contributions}

\revision{
As learner agency has become central to discussions of how increasingly capable AI systems and educational technologies mediate, evaluate, and govern learning, its importance is widely acknowledged (e.g., \cite{darvishi2024impact, brod2026agency, viberg2025fostering}). Yet, despite its growing rhetorical prominence, agency remains weakly theorized and inconsistently operationalized. Our review of the Learning at Scale literature (Section \ref{sec:lit-review}) shows that learner agency is frequently invoked indirectly through adjacent concepts or proxies such as engagement, personalization, self-regulation, or help-seeking, often without explicit definitions or stated assumptions about decision-making authority. As a result, claims about whether learning systems support or undermine learner agency are difficult to compare, reproduce, or evaluate across extended learning trajectories and settings.
}

\revision{
This position paper treats learner agency as a core priority for the next phase of investigating learning and teaching at scale. We contribute \textbf{(1)} a synthesis of four interrelated challenges that constrain the study and design of learner agency at scale, and \textbf{(2)} the \AAFshort{} for researchers aiming to study and foster learner agency. The \AAFshort{} clarifies when and where agency is enacted, who participates in decisions, how choices are structured, and over what timescales their consequences unfold. Together, these contributions aim to provide scholars with shared conceptual and analytic tools for systematically reasoning about learner agency in AI-mediated, large-scale learning environments.
}

\section{Comparative Literature Review}
\label{sec:lit-review}

We searched the ACM Digital Library for Learning at Scale conference papers only, using agency-related search terms, and did not aim for a comprehensive review across learning sciences, HCI, or AI ethics.
\revisionthree{
We deliberately limit this review to the proceedings of the ACM Learning at Scale conference.
This focus allows us to identify recurring themes that arise specifically in scaled and automated contexts.
We clarify where the field lacks shared assumptions about decision authority, evidence, and evaluation horizons.
Grounding the analysis in this field-specific gap enables us to articulate concrete research problems and to position the \AAFshort{} as a tool for advancing future work within (and in dialogue with) adjacent literatures.
}
Our search targeted papers that included the term \emph{agency} in the title or abstract. The only prior Learning at Scale paper with human learner agency explicitly in the title is Xie et al.~\cite{xie2020effect}. In this work, the authors conceptualized agency as learners' ability to take actions aligned with their learning goals. They operationalized it through system design by manipulating learners' degree of choice and the availability of proximal information. They evaluated a self-directed Python learning environment with three conditions: informed high-agency, uninformed high-agency, and informed low-agency. While higher agency and information were associated with increased engagement and motivation, they did not lead to improved learning outcomes, which the authors attributed to the additional decision-making burden placed on learners. All other papers focused on artificial agents (e.g., LLM agents) rather than human learner agency.

We then broadened our search to include related constructs and terminology commonly used to study agency in large-scale learning environments. Drawing on Brod's theoretical review of learner agency \cite{brod2026agency}, we derived four additional search terms, since the construct of agency is often discussed through the following different labels: \textbf{Learner control}, \textbf{Self-directed learning}, \textbf{Metacognitive control}, and \textbf{Help-seeking}. Under \textbf{Learner control}, we identified one additional Learning at Scale paper \cite{wan2017one}. However, this work primarily focuses on system-driven personalization by matching tutoring interventions to students, rather than granting learners meaningful control over instructional decisions. Several papers aligned with \textbf{Self-directed learning}, including work on student-driven study planning using LLMs \cite{chun2025planglow}, a qualitative study of teacher perspectives on supporting self-regulated learning skills in an at-scale tutoring system \cite{brezack2025teacher}, and a generalizable approach to automatically coding log data to identify educator support roles and student actions consistent with contemporary conceptualizations of 21st-century learning \cite{nacu2016beyond}. Our search returned no results explicitly addressing \textbf{Metacognitive control} in large-scale learning environments. In contrast, multiple studies addressed \textbf{Help-seeking} behaviors, including an experimental study of feedback designs in a community Q\&A system \cite{frens2018supporting}, work grounded in self-determination theory that examined how structured support strategies in MOOCs influenced learner retention and self-regulation behaviors such as help seeking \cite{chen2014facilitating}, an analysis of instructor help-seeking in a Canvas community forum \cite{ramesh2021setting}, and an evaluation of crowdsourced teacher-generated tutoring content in ASSISTments that found uneven effectiveness across contributors and limited suitability for personalization based on student knowledge alone \cite{prihar2021toward}.

\revisiontwo{
Across the ten studies identified, engagement with learner agency remains limited and fragmented. Agency is rarely examined directly. Instead, it appears through proxy constructs such as learner control, self-directed learning, self-regulation, personalization, and help-seeking. Few studies explicitly theorize agency, and none operationalize it as a distinct construct. Existing research also overlooks shifting control in human–AI learning contexts where agency may be shared or constrained. This highlights the need to clarify what learner agency means in AI-mediated learning and how it can be intentionally supported, measured, and evaluated at scale through a unifying framework applicable across K–12 and higher education.}

\section{Four Challenges for Learner Agency at Scale}

The following four challenges articulate key points at which existing theories, methods, and design practices break down when applied to large-scale learning environments. Together, they surface foundational obstacles that must be addressed to study, design for, and sustain learner agency at scale. We structure each challenge around a central claim that identifies a recurring limitation in current research and practice in teaching and learning platforms that operate at scale, followed by an illustrative example that grounds the challenge in concrete system designs or empirical findings. Each challenge concludes by outlining how an alternative research or design orientation could address the limitation, highlighting opportunities for researchers to more explicitly account for learner agency in large-scale, technology- and AI-mediated environments.

\subsection{Challenge 1: Conceptual Ambiguity}

\textbf{Claim.} Learner agency is invoked in AI in education, learning analytics, and learning at scale research (e.g., \cite{mouta2025agency, viberg2025fostering, xie2020effect}), yet it remains conceptually underspecified and inconsistently defined across disciplines and system designs. Human agency is commonly described as the capacity to act intentionally, make informed choices, exert meaningful control, and influence outcomes \cite{brod2026agency}. However, sociocultural \cite{vygotsky1977development}, self-determination \cite{deci2012self}, and social cognitive perspectives \cite{bandura2006toward} conceptualize agency in distinct ways, emphasizing relational dynamics, autonomy, self-regulation, and contextual constraints. In large-scale AI-mediated learning environments, this conceptual diversity often collapses into narrow operationalizations where agency is inferred from observable behavior rather than grounded in coherent theoretical accounts. As AI systems increasingly shape learning pathways, feedback, and decision-making in institutional and classroom contexts \cite{floridi2024ai,pardos2026oecd}, the absence of a shared definition risks obscuring whether learners are exercising meaningful control or merely responding to automated structures.

A further source of ambiguity concerns whether agency should be understood as an individual capacity or a collective and infrastructural phenomenon. While educational theory often locates agency within learners \cite{brod2026agency,tay2025role,winstone2017supporting}, research on AI-mediated socio-technical systems highlights that agency is distributed across learners, educators, institutions, and technological infrastructures \cite{suchman2007human}. In practice, adaptive tutoring systems, recommendation engines, and automated feedback tools can redistribute decision-making authority in ways that blur the boundary between human intention and system delegation \cite{natarajan2025human}. Without explicitly theorizing how agency is shared, constrained, or transferred across these actors, it is difficult to assess who is actually shaping learning outcomes.

\revisionthree{To avoid relabeling other constructs, we adopt a definitional boundary: for this framework, agency concerns the locus of \emph{decision-making authority} over learning-relevant decisions (see Section \ref{sec:framework} describing the \AAFshort{}). This authority describes \textit{who} can initiate, veto, or override an action (Step 2; \AAFshort{}). Decision-making also encompasses which options students choose (Step 1; \AAFshort{}) and with what information and choice architecture design (Step 3; \AAFshort{}). This distinguishes agency as described in the \AAFshort{} from learner control (which does not specify stakeholders and choice architecture), self-regulated learning (processes that may be under or outside learner control), autonomy support (an environmental condition that enables agency but is not agency itself), human-in-the-loop and contestability (design features that affect but are not equivalent to where authority resides), and engagement (observable participation that need not entail control or consequential choice). The \AAFshort{} operationalizes agency by requiring researchers to specify which decisions and who holds authority, so that claims are about allocation of control and decision-making.}

The conflation of short-term expressions of agency with long-term developmental trajectories further compounds conceptual ambiguity. Many educational technologies infer agency from immediate choices, engagement patterns, or responsiveness to prompts (e.g., \cite{winstone2017supporting,tay2025role,borchers2025learner}), yet agency also unfolds over time through identity formation, epistemic judgment, metacognitive growth, and sustained ownership of learning goals \cite{kizilcec2023pipelines,bandura2006toward,nasir2006exploring}. Treating momentary action as equivalent to enduring agency risks overstating the extent to which AI-mediated systems support learner agency.

Finally, agency is frequently confused with system automation or efficiency. As AI systems take on greater responsibility for planning, assessment, and cognitive scaffolding, improvements in personalization or performance are sometimes interpreted as gains in learner empowerment. However, delegating cognitive or metacognitive work to AI does not necessarily strengthen learners' capacity for intentional action. Instead, it may reduce opportunities for reflection, decision-making, and meaningful control \cite{salomon1991partners,fan2025beware}. Moreover, research and system design often embed implicit normative assumptions about what learners should value or optimize, such as speed, completion, or standardized performance, without making these commitments explicit \cite{williamson2019policy}.

\textbf{Example and Elaboration.} These ambiguities become visible in AI-mediated learning systems that present learners with adaptive pathways or automated recommendations. Such systems (e.g., \cite{pardos2023oatutor}) can expand access to personalized support. Yet, they also can shift researcher attention from supporting learners to optimizing algorithms when deciding what to study and how to evaluate success (e.g., via engagement metrics). Learners may appear active and self-directed while system constraints tightly bind their options. In these cases, it is unclear whether learners are exercising agency or merely navigating predefined choices shaped by institutional and technological priorities.

\textbf{An Alternative Path for Learning at Scale Research.} Addressing conceptual ambiguity requires treating learner agency as a theoretically grounded construct. Learning at Scale research should clarify what counts as agency in automated and AI-mediated environments by distinguishing between individual and collective agency, short-term action and long-term development, and human control and system delegation. Rather than concluding agency from engagement or performance, researchers should explicitly model where decisions reside, what alternatives learners can access, and how their capacity to shape learning evolves. Establishing shared conceptual foundations is essential for measuring, designing for, and sustaining learner agency in large-scale learning ecosystems.

\subsection{Challenge 2: Measuring Agency at Scale}

\textbf{Claim.} Engagement occupies a central place in learning at scale research because it is readily observable and predictive of short-term learning outcomes \cite{koedinger2015learning}. Clicks, time-on-task, persistence, responsiveness to prompts, and completion rates are often treated as indicators of successful learning experiences. This emphasis is understandable: engagement is measurable at scale, amenable to experimentation, and actionable for system design. However, engagement is an insufficient and potentially misleading proxy for learning \cite{gavsevic2015let}. Classic Learning at Scale results, such as Koedinger et al.'s ``Learning is Not a Spectator Sport,'' demonstrate that active responding and practice lead to better learning than passive exposure to educational contexts such as videos \cite{koedinger2015learning}. Yet active engagement should not be conflated with agency. Learners can be highly active by clicking, responding, and persisting while exercising little intentional control over what they are doing or why. In contrast, lack of engagement and delays in completing assignments can be strategic and even beneficial for some students, as shown in recent procrastination research \cite{kim2025not}. In many at-scale learning systems that scaffold learner problem solving, engagement is produced through adaptive and personalized algorithms that decide over the sequence and nature of learning activities \cite{pardos2023oatutor,heffernan2014assistments}. Only sometimes are teachers \cite{heffernan2014assistments} or students directly involved in selecting activities, for example, in open-learner models where students register preferences for what skill to practice \cite{borchers2025learner}. \revisiontwo{A complication is that evaluation of agency may emphasize perceived control (e.g., via surveys) or observable behavior (e.g., log data), and the two need not align.}

\textbf{Example and Elaboration.} Another example includes pathway-level recommendation and pacing systems that optimize engagement and predicted success across courses or units. In many large-scale programs, learners are guided by default sequences, recommended next courses, or adaptive pacing rules designed to increase persistence and completion rates \cite{kizilcec2023pipelines}. These systems can be effective at reducing attrition, but they also shift planning and foresight from learners to institutional or algorithmic logic. Over time, learners may follow recommended pathways without evaluating trade-offs, considering alternatives, or reflecting on how different choices align with their longer-term goals \cite{davis2016gauging}. Even when systems are designed to incorporate learner input \cite{shao2021degree,reinhard2024one,pardos2017enabling}, such as degree planning tools that allow students to specify goals and receive optimized course sequences \cite{shao2021degree}, decision authority remains largely algorithmic, and the consequences of those recommendations are often opaque. Engagement and completion remain high, but opportunities for learners to practice agency through consequential planning and ownership of academic trajectories may be reduced once algorithmic support is reduced. Even simple instructional design choices, such as the order of presentation of course materials \cite{davis2016gauging, kizilcec2015instructor} or the timing of deadlines \cite{tao2025investigating}, can influence learner behavior and, over time, erode their agency.

A related pattern is emerging in AI-supported feedback and writing environments. Systems that provide individualized, adaptive feedback have been shown across experimental studies to produce medium positive effects on subsequent writing performance \cite{fleckenstein2023automated}. However, improved performance does not necessarily imply that learners are independently evaluating feedback or engaging in self-monitoring. When learners readily accept system-generated suggestions, engagement metrics may increase while opportunities to practice evaluative judgment decline.

\textbf{An Alternative Path for Learning at Scale Research.}
A productive shift beyond the field's focus on engagement is to examine when engagement reflects intentional, informed, and consequential learner action. Rather than asking whether learners are active or persistent, agency research should ask who makes decisions, what alternatives exist, and whether learners can develop control over their learning trajectories over time \cite{schwartz2013measuring}. This does not mean abandoning engagement as a useful signal, but situating it within a broader account of agency. Without this distinction, large-scale systems may optimize observable participation while obscuring whether learners are developing the capacity to plan, choose, and act in line with their goals. Recognizing agency as distinct from engagement is therefore essential for responsible interpretation, evaluation, and design of learning at scale.

\subsection{Challenge 3: Agency and Efficiency}

\textbf{Claim.} Large-scale learning environments are built to deliver instruction reliably across diverse learners, contexts, and institutional constraints. Achieving scale typically requires \textit{standardization}: stable content sequences, consistent rules, predictable pacing structures, and automated workflows that reduce instructor burden and enable evaluation and maintenance. Learner agency, however, is expressed through \textit{deliberation}, \textit{variation}, and \textit{time}: learners need opportunities to set goals, explore alternatives, reflect on trade-offs, and adapt plans in response to changing circumstances. This creates a persistent design tension for scaled learning platforms: the mechanisms that make systems efficient, scalable, and comparable (e.g., defaults, recommended pathways, uniform pacing) often reduce the space for meaningful learner control, while designs that expand choice and flexibility can impose cognitive and logistical costs that undermine learning, equity, and implementation fidelity.

A central trade-off concerns \textit{choice and cognitive load due to choice complexity}. Providing learners with more options is often treated as synonymous with supporting agency. Yet, more choice can increase decision burden and reduce performance, especially for novices or learners with fewer supports \cite{xie2020effect}. Indeed, learners often do not prefer high levels of choice due to the complexity and the relinquishing of some control to algorithms when selecting learning tasks \cite{borchers2025learner}. At scale, even small increases in decision complexity (e.g., selecting among problem types or courses) can translate into overwhelm or suboptimal decisions (e.g., watching videos rather than engaging in problem-solving in MOOCs \cite{koedinger2015learning}). Conversely, strong guidance can reduce cognitive load and increase completion, but can also shift planning and evaluative judgment away from learners. The result is a design dilemma: systems can optimize for ease and throughput by narrowing choice, or provide richer opportunities for agency while risking overload and uneven uptake.

A second trade-off concerns \textit{autonomy and guidance}. Scaffolds such as recommendations and automated feedback can help learners progress. Still, they can also shape what learners perceive as viable or worthwhile, especially when system suggestions are framed as personalized or authoritative \cite{shao2021degree,reinhard2024one,davis2017follow,RafieianZuo2024Personalization}. These design patterns can improve efficiency and reduce attrition. Still, they can also reduce learners' opportunity to practice goal setting, monitoring, and reflective choice, which are core components of agency \cite{brod2026agency}. In other words, systems may help learners \textit{get through} learning while weakening their capacity to \textit{steer} it.

A third trade-off arises between \textit{equity goals and personalization}. Personalization can support agency by aligning activities with learners' goals, identities, and contexts. Still, personalization mechanisms can also introduce differential experiences that are difficult to audit and may reproduce structural inequities. For example, systems that optimize for predicted success may systematically steer some learners toward ``safer'' pathways that limit access to challenging material or advanced tracks, even when learners would value exploration or long-term growth \cite{harrison2022should,malik2025transcripts,dalberg2024major}. These forms of differential decision-making may be grounded in relatively stable personality traits and dispositions, which have received little attention in educational technology research \cite{borchers2026toward}. Equity-oriented design may require standardization and guardrails (e.g., minimum exposure to key content, constraints against tracking), which can conflict with individualized pathways that appear more agentic. Learning at Scale systems must thus negotiate whether agency should be expressed through student choice, through equitable access to consequential opportunities, or through a balance of both.

Finally, Learning at Scale systems face trade-offs between \textit{implementation fidelity and learner adaptation}. At scale, interventions often require consistency to be evaluated and deployed across courses, instructors, or institutions. Yet learners' agency is inherently adaptive: learners reinterpret tasks, create workarounds, resist or repurpose prompts, and pursue goals that differ from designer assumptions \cite{baker2008students}. Designs that enforce fidelity through strict policies can stabilize evaluation and operations, but can also suppress the very adaptations through which learners exercise agency. Conversely, designs that allow extensive adaptation can undermine comparability across populations. This tension becomes especially salient in AI-mediated environments, where learners can dynamically negotiate, bypass, or outsource parts of the learning process via external tools (see also Challenge 4).

\textbf{Example and Elaboration.} These trade-offs become visible in systems that offer learners more control over sequencing or pacing while attempting to preserve learning effectiveness. For instance, environments that increase learner control by allowing students to choose what to study next can improve motivation and engagement, but may impose additional decision-making costs \cite{xie2020effect,borchers2025learner}. Similarly, large-scale tutoring systems often rely on adaptive problem selection to maximize learning efficiency, reducing the need for learners to plan and monitor, but also limiting opportunities for learners to choose practice goals \cite{heffernan2014assistments,pardos2023oatutor,borchers2025learner}. In both cases, the design question is not whether to support learners or to give choices, but how to allocate decision authority so that learners can exercise meaningful control without being overburdened or unevenly advantaged.

\textbf{An Alternative Path for Learning at Scale Research.} Taken together, these challenges suggest a new Learning at Scale research goal: to study and design \textit{bounded agency}, systems that preserve scalable structure while creating targeted, developmentally appropriate decision points where learner control is meaningful and learnable. This implies designing choice architectures that (a) limit options to those that are interpretable and consequential, (b) provide transparent trade-offs and reversible decisions when possible, and (c) scaffold learners' ability to make choices over time rather than assuming agency is immediately available (cf. choice-based assessments for measuring learning \cite{schwartz2013measuring}). It also implies evaluating agency not as the sheer availability of options, but as learners' ability to make informed, effective, and goal-aligned decisions under real constraints, including time, workload, limited resources, and institutional policies. Developing measures of bounded agency requires models that capture decision-making under constraint. Computational models from cognitive psychology, which formalize how people make decisions with limited time, information, and cognitive resources, offer a promising foundation for modeling and estimating agency across learning contexts \cite{lieder2020resource}.

Crucially, these trade-offs should be made explicit and evaluated as design commitments. The \AAFshort{} (Figure~\ref{fig:decision-framework-agency}) provides one way to operationalize this shift: by specifying which decisions are in view, who has authority, how choices are structured, and over what horizons consequences unfold, researchers can compare systems based on how they balance efficiency, equity, and meaningful learner control. Making these trade-offs visible is a necessary step toward designing large-scale learning environments that are not only effective and scalable, but also agency-preserving over time.

\subsection{Challenge 4: AI-Mediated Agency}

\textbf{Claim.} In large-scale learning environments, learner agency is increasingly shaped by algorithmic and AI-mediated systems that recommend, prioritize, and constrain action. Adaptive tutors, recommender systems, and generative AI now participate directly in planning, pacing, assessment, and problem solving. As a result, agency is no longer exercised solely by learners but is redistributed across learners, designers, institutions, and technical infrastructures. While this redistribution enables scale and personalization, it also introduces a fundamental challenge for Learning at Scale research: it becomes unclear who is responsible for decisions that shape learners' opportunities, trajectories, and outcomes, and on what basis those decisions should be evaluated.

In many AI-mediated learning systems, planning and decision-making are partially delegated to algorithms that operate on predicted success, engagement, or efficiency \cite{kizilcec2023pipelines,xu2023convincing}. These systems determine which tasks are shown, which hints are offered, when learners advance, and which pathways are encouraged or foreclosed. Learners retain the ability to act within these systems, but system-generated defaults, rankings, and constraints condition their actions. Agency is thus exercised through interaction with algorithmic logic (with differing levels of interpretability \cite{swamy2022evaluating,xu2023convincing,harrison2022should}), blurring the distinction between learner intention and system guidance.

This redistribution of agency is further complicated by learners' adaptive responses to system behavior. Over time, learners learn how systems work and adjust their actions accordingly, following recommendations to minimize effort, gaming incentives, or deferring judgment to AI-generated feedback \cite{fan2025beware,salomon1991partners}. What appears as learner choice and agency may therefore reflect strategic compliance with system logic rather than independent deliberation. At scale, such adaptations can systematically reshape how agency is expressed, often in ways that align with system optimization goals rather than learners' longer-term interests. \revisionthree{The extent to which automation and such adaptations attenuate agency in practice remains an open empirical question. The \AAFshort{} makes it testable by requiring explicit specification of who holds authority before and after system use.} A fundamental challenge for Learning at Scale, therefore, is to study \textit{how} learner-AI decision-making evolves and \textit{what} different forms of learner-AI interaction entail.

A central difficulty is that algorithmic systems typically optimize tractable proxies of agency, such as engagement, persistence, or short-term performance (see Section \ref{sec:lit-review}). These proxies are necessary for operationalization at scale, but they risk substituting measurable behavior for meaningful control. Systems may appear to ``support agency'' by offering choices or personalization while still constraining which alternatives are visible, reversible, or consequential. When agency is inferred solely from these proxies, Learning at Scale research risks overstating learner agency while underexamining how decision authority is actually allocated.

Finally, accountability for agency-related outcomes becomes unclear when decisions are distributed across learners, educators, institutions, and algorithmic systems. If a learner follows a recommended pathway that later limits course options, career trajectories, or time to degree, responsibility becomes difficult to assign: did the learner choose the path, did the model shape the choice, did the designer constrain the alternatives, or did institutional policy define the available options? Similar questions arise when students accept automated feedback, follow system-generated pacing, or rely on recommended next actions that shape their progress over time. In these cases, outcomes emerge from a chain of human-AI decisions rather than a single, clearly attributable choice. Few Learning at Scale evaluations make these allocations explicit, focusing instead on aggregate outcomes such as completion, engagement, motivation, grades, or persistence \cite{koedinger2023astonishing,baker2008students}. Yet as AI systems increasingly influence access, progression, and evaluation, questions of responsibility and contestability become unavoidable. From both research and regulatory perspectives, learning systems must support the ability to understand how decisions were made, who influenced them, and how they can be challenged or revised \cite{kay2023scrutable}.

\textbf{Example and Elaboration.} The described issues are visible in AI-supported planning and recommendation systems used for course selection, pacing, or tutoring. Degree planners and adaptive tutors may allow learners to specify goals while algorithmically generating optimized sequences \cite{shao2021degree,pardos2023oatutor}. Although learners appear to be in control, key decisions, including what trade-offs are considered, which risks are acceptable, and which paths are excluded, are embedded in system logic and often opaque. Learners may accept recommendations as authoritative, especially when they are framed as data-driven or personalized, even when those recommendations encode institutional priorities such as efficiency or throughput \cite{RafieianZuo2024Personalization,hariyanto2025artificial}. In such cases, agency is not removed but reconfigured, and its consequences may only become visible once options are no longer reversible.

\textbf{An Alternative Path for Learning at Scale Research.} Addressing algorithmic and AI-mediated agency requires Learning at Scale research to explicitly measure and examine how decision-making authority is distributed and governed. Researchers can analyze which decisions are automated, which remain human, and how learners can understand, contest, or override system recommendations. This includes making visible the value systems being optimized, the evidence they rely on, and who bears responsibility for downstream consequences.

Importantly, this challenge connects directly to emerging regulatory and governance pressures. Frameworks such as GDPR and the EU AI Act place limits on automated decision-making in high-risk settings, including education, and emphasize requirements for human autonomy, transparency, and accountability \cite{voigt2017eu,veale2021demystifying}. For research and design, this implies specifying which decisions are fully automated versus subject to human review, ensuring that learners or educators can obtain meaningful information about how decisions were made and contest outcomes \cite{hutt2023right}. These regulations raise practical questions about how learner-facing AI systems should be designed and evaluated when they shape course pathways, feedback, or access to support. Learning at Scale research is well-positioned to address these questions by providing empirical evidence on how agency is allocated between learners, instructors, and systems in real deployments, and how this allocation changes over time. Doing so requires evaluation approaches that go beyond predictive or performance metrics to make shifts in decision authority visible, open to challenge, and subject to ongoing oversight.

\section{The Agency Allocation Framework (AAF)}
\label{sec:framework}

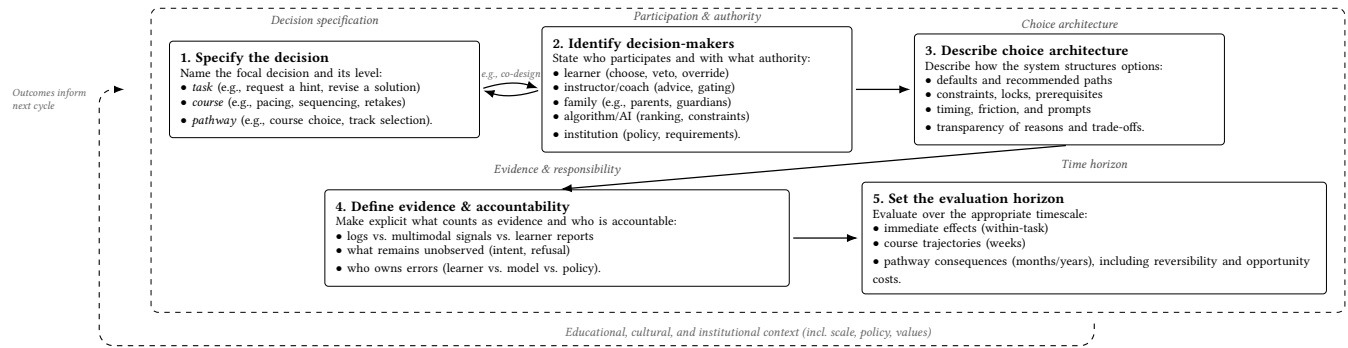
\begin{figure*}[htpb]
\centering
\resizebox{\textwidth}{!}{%
\begin{tikzpicture}[
  font=\small,
  >={Latex[length=2mm]},
  focus/.style={
    draw,
    rounded corners=2pt,
    fill=black!3,
    align=left,
    inner sep=7pt,
    text width=120mm
  },
  stageTop/.style={
    draw,
    rounded corners=2pt,
    align=left,
    inner sep=6pt,
    text width=56mm
  },
  stageRight/.style={
    draw,
    rounded corners=2pt,
    align=left,
    inner sep=6pt,
    text width=56mm
  },
  stageNarrow/.style={
    draw,
    rounded corners=2pt,
    align=left,
    inner sep=6pt,
    text width=56mm
  },
  stageBottomL/.style={
    draw,
    rounded corners=2pt,
    align=left,
    inner sep=6pt,
    text width=86mm
  },
  stageBottomR/.style={
    draw,
    rounded corners=2pt,
    align=left,
    inner sep=6pt,
    text width=86mm
  },
  context/.style={
  draw,
  dashed,
  rounded corners=4pt,
  inner sep=10pt,
  fit=#1,
  label={[font=\footnotesize\itshape, text=black!60,yshift=-5pt]below:Educational, cultural, and institutional context (incl. scale, policy, values)}
  },
  flow/.style={-Latex, line width=0.7pt, shorten <=2pt, shorten >=2pt},
  flowBoth/.style={Latex-Latex, line width=0.7pt, shorten <=2pt, shorten >=2pt},
  flowLoop/.style={-Latex, line width=0.6pt, shorten <=2pt, shorten >=2pt, dashed},
  tag/.style={font=\footnotesize\itshape, text=black!70}
]

\node[stageTop, anchor=north] (which) {\textbf{1. Specify the decision}\\
\footnotesize
Name the focal decision and its level:\\
$\bullet$ \emph{task} (e.g., request a hint, revise a solution)\\
$\bullet$ \emph{course} (e.g., pacing, sequencing, retakes)\\
$\bullet$ \emph{pathway} (e.g., course choice, track selection).};

\node[stageRight, right=12mm of which] (who) {\textbf{2. Identify decision-makers}\\
\footnotesize
State who participates and with what authority:\\
$\bullet$ learner (choose, veto, override)\\
$\bullet$ instructor/coach (advice, gating)\\
$\bullet$ family (e.g., parents, guardians)\\
$\bullet$ algorithm/AI (ranking, constraints)\\
$\bullet$ institution (policy, requirements).};

\node[stageNarrow, right=12mm of who] (how) {\textbf{3. Describe choice architecture}\\
\footnotesize
Describe how the system structures options:\\
$\bullet$ defaults and recommended paths\\
$\bullet$ constraints, locks, prerequisites\\
$\bullet$ timing, friction, and prompts\\
$\bullet$ transparency of reasons and trade-offs.};

\node[stageBottomL, below=10mm of which, anchor=north west] (evidence) {\textbf{4. Define evidence \& accountability}\\
\footnotesize
Make explicit what counts as evidence and who is accountable:\\
$\bullet$ logs vs. multimodal signals vs. learner reports\\
$\bullet$ what remains unobserved (intent, refusal)\\
$\bullet$ who owns errors (learner vs. model vs. policy).};

\node[stageBottomR, right=14mm of evidence] (horizon) {\textbf{5. Set the evaluation horizon}\\
\footnotesize
Evaluate over the appropriate timescale:\\
$\bullet$ immediate effects (within-task)\\
$\bullet$ course trajectories (weeks)\\
$\bullet$ pathway consequences (months/years), including reversibility and opportunity costs.};

\node[tag, above=1.5mm of which] {Decision specification};
\node[tag, above=-0.5mm of who] {Participation \& authority};
\node[tag, above=0mm of how] {Choice architecture};
\node[tag, above=1.5mm of evidence] {Evidence \& responsibility};
\node[tag, above=1.5mm of horizon] {Time horizon};

\draw[flow] (which.east) to[bend left=18] (who.west);
\draw[flow] (who.west) to[bend left=18]
  node[midway, sloped, above=5pt, font=\scriptsize\itshape, text=black!60]
  {e.g., co-design}
(which.east);
\draw[flow] (who.east) -- (how.west);

\draw[flow] (how.south) -- (evidence.north);
\draw[flow] (evidence.east) -- (horizon.west);

\node[context={(which)(who)(how)(evidence)(horizon)}] (ctx) {};

\draw[flowLoop, rounded corners=8pt]
  ([yshift=-5mm]horizon.south) -- ++(0,-5mm)
  -| ([xshift=-10mm]ctx.west)
  |- ([xshift=-8mm]which.west);

\node[font=\scriptsize\itshape, text=black!60, align=left, anchor=east]
  at ([xshift=-15mm,yshift=-2mm]which.west)
  {Outcomes inform\\next cycle};

\end{tikzpicture}%
}
\caption{The AAF treats agency as enacted through a focal decision point and proceeds through five steps. The link between specifying the decision and identifying decision-makers can be bidirectional (e.g., when co-designing with stakeholders), and outcomes can inform the next cycle of design and experimentation.}
\label{fig:decision-framework-agency}
\end{figure*}

Large-scale learning environments make learner agency both consequential and difficult to observe. At scale, instructional decisions that were once negotiated between learners and instructors are increasingly standardized, automated, or delegated to algorithmic systems. As a result, agency can be easily overlooked, because opportunities to act intentionally are embedded in system-level decision structures, not just expressed through overt choice. Figure~\ref{fig:decision-framework-agency} introduces the \AAFshort{} designed to make these structures explicit.

The \AAFshort{} centers on a \textbf{Decision Point: where agency is enacted}. Learner agency in technology-mediated, large-scale environments is enacted through \textbf{decisions} that shape learning trajectories: what to do next, when to progress, what support to accept, and which goals or pathways remain viable over time. In at-scale settings, these decisions are rarely made by learners alone. Decision authority is distributed across learners, educators, families, institutions, and AI or algorithmic infrastructures that rank options, enforce constraints, and define defaults \cite{roll2018learning,kizilcec2023pipelines,suchman2007human}. As Figure~\ref{fig:decision-framework-agency} illustrates, the \AAFshort{} specifies what decision is in view, who participates, how it is structured, what evidence is used to justify it, and over what time horizon consequences unfold, providing a shared language for examining this distribution of responsibility and control.

\paragraph{Relation to SRL, SDT, and existing agency theories.}
\revisionthree{The \AAFshort{} isolates two components central to agency: \emph{control} (who can initiate, veto, or override) and \emph{decision-making} (who chooses among options and with what information) \cite{brod2026agency}. From broader theoretical perspectives such as SRL and SDT, we operationalize by asking where these components reside in a given system.} The \AAFshort{} is intentionally complementary to established theories such as Self-Regulated Learning (SRL) and Self-Determination Theory (SDT). SRL models describe how learners plan, monitor, and regulate their learning when those processes are under their control, while SDT emphasizes the motivational importance of autonomy, competence, and relatedness. Neither was developed for environments in which planning, monitoring, and choice are partially delegated to automated systems, nor for settings in which learners' opportunities to self-regulate are shaped by institutional and algorithmic constraints. The \AAFshort{} does not redefine agency or learner motivation. It specifies the \emph{decision context} under which SRL and autonomy can plausibly occur at scale. \revisionthree{This sharpens the contrast with engagement: engagement does not require learner control or consequential decision-making, whereas the framework treats agency as the allocation of exactly those two components between system and learner.}

\paragraph{Overview of the five steps.}
Figure~\ref{fig:decision-framework-agency} presents five analytic steps. The first two steps prompt researchers to define the forms of learner decisions they want to study and the level of analysis and stakeholders relevant to their study. The link between steps 1 and 2 can be bidirectional when researchers or designers co-design with stakeholders: who is involved may shape how the decision is specified, and vice versa, depending on the maturity of the system under study. Step 3 (choice architecture) asks researchers to describe how the system structures the options available to learners: defaults and recommended paths, constraints and prerequisites, timing and friction, and the transparency of reasons and trade-offs. Step 4 (evidence and accountability) requires researchers to state what counts as evidence for the decision (e.g., logs, multimodal signals, learner reports), what remains unobserved (e.g., intent or refusal), and who is accountable for errors or outcomes. Step 5 (evaluation horizon) asks researchers to set the timescale over which consequences are evaluated, from immediate or within-task effects to course-level trajectories to pathway consequences over months or years, including reversibility and opportunity costs. The dashed arrow in the figure indicates that outcomes can inform the next cycle of design and experimentation, thereby supporting iterative use of the framework. Step order can vary depending on whether researchers are analyzing an existing system or co-designing a new one.

\paragraph{Who makes decisions at scale.}
A common source of ambiguity in agency discussions is the assumption that decisions have a single decision-maker. In practice, decisions about learning at scale are layered. Institutions define policies and requirements. Platforms translate these into interfaces and workflows. Algorithms generate rankings, defaults, or triggers. Educators may approve, interpret, or override system outputs. Families (e.g., parents or guardians) may influence access, pacing, or goals in many learning settings. Learners act within this structured space, sometimes choosing, sometimes complying, sometimes resisting, and sometimes deferring judgment to the system. Agency, therefore, cannot be assessed by the presence or absence of choice alone. It depends on which decisions are available, which are consequential, and what forms of authority learners retain relative to system defaults and constraints.

\paragraph{Behavioral grounding and its advantages.}
The \AAFshort{} deliberately leans on decisions and system structures because these are observable and measurable at scale. Behavioral and multimodal data provide access to when decisions occur, what options were available, and how learners responded under specific constraints. This makes the \AAFshort{} compatible with dominant Learning at Scale methodologies. At the same time, the \AAFshort{} avoids behavioral reductionism. Observable actions are treated as evidence about how a decision context is navigated, not as direct indicators of intention. By prompting researchers to state what remains unobserved and how responsibility is assigned (Step 4), the \AAFshort{} supports cautious \emph{interpretation} of agency rather than concluding agency from engagement alone.

\paragraph{Context, culture, and scale.}
As the figure indicates, the five steps are situated within \textbf{educational, cultural, and institutional context} (including scale, policy, and values). Learner context can be a conditioning force that shapes how agency is enacted and enabled in scaled-up learning systems. Cultural norms, educational values, institutional missions, and learner characteristics influence which options are legible, which risks are culturally acceptable, and which decisions are possible. At scale, these contextual factors can interact with standardization and automation, often producing differential effects across learners \cite{borchers2026toward}. The \AAFshort{} prompts researchers to examine how the same system design can support agency in one context while constraining it in another, helping to explain unintended consequences that only emerge under large-scale deployment.

\section{Sample Application}

To illustrate the \AAFshort{}, we apply it to an at-scale tutoring context, \revisiontwo{showing how researchers can operationalize its five steps in practice: specifying the focal decision, identifying who participates and with what authority, describing the choice architecture, defining evidence and accountability, and setting the evaluation horizon.} Large-scale tutoring systems, including intelligent tutoring systems and increasingly LLM-mediated dialogue tutors, provide a useful test case because they combine fine-grained interaction data with substantial automation over instructional decisions such as task selection, sequencing, and feedback \cite{koedinger2015learning,borchers2025learner}. While designed to support problem solving and self-regulation, these systems can also shift planning and regulatory work away from learners. We therefore apply the framework to a recurring tutoring decision: how learners respond to difficulty during problem solving.

\revisionthree{Practically, the \AAFshort{} promotes a preregistration-like commitment: researchers specify the focal decision, decision-makers, choice architecture, evidence, and evaluation horizon in advance (whether before data collection, analysis, or when auditing existing systems). The aim is to make assumptions explicit and reviewable. The framework remains flexible and iterative (e.g., supporting co-design between Steps 1 and 2). We demonstrate its use with a tutoring-system example, a core case of large-scale learning \cite{koedinger2015learning,koedinger2023astonishing}.}

\textbf{Step 1: Specify the decision.}
\revisiontwo{Agency can be exercised and enacted in several forms of choice \cite{brod2026agency}, ranging from course and major selection in higher education to micro-level interactions in adaptive learning systems \cite{fischer2020mining}. Here, we focus on the latter, since learning interfaces are increasingly embedded with AI and analyzable through design and log data lenses \cite{darvishi2024impact,borchers2026brief}. Based on these considerations, the first step in the \AAFshort{} is for the researcher to define \textit{what} the decision is that the learner can engage in and \textit{how} it manifests and is observable. In our example, the focal decision is how a learner proceeds when they encounter difficulty during problem solving, a \emph{task-level} decision and key theoretical foundation of persistence \cite{tinto2017reflections}.} In at-scale tutoring, the options visible to the learner are typically limited to in-system actions (e.g., request a hint, attempt another step, ask a follow-up in dialogue, wait for system intervention, or disengage); \revisiontwo{out-of-system options such as consulting a teacher or family member, taking a break, or checking a textbook are either unavailable or invisible in the system's decision space, which the \AAFshort{} calls the ``choice architecture'' (Step 3).} While this decision appears local \revisiontwo{and limited}, it is consequential because it recurs frequently and shapes learners' emerging help-seeking strategies and reliance on system support over time. At scale, small differences in how this decision is structured can accumulate into stable patterns of dependency or self-regulation.

\textbf{Step 2: Identify decision-makers.}
\revisiontwo{Once the decision is specified, the second step asks researchers to identify \textit{who} participates in it and with what form of authority (e.g., to choose, veto, or override).} For instance, decisions about help-seeking in tutoring contexts are jointly produced by multiple actors. Learners initiate actions, but the tutoring system determines when hints are available, how many attempts are permitted, and whether unsolicited guidance is triggered. In LLM-mediated tutoring, the system may further decide whether to respond with probing questions or scaffolding and explanations. Instructors or institutional policies may indirectly shape this decision by enforcing curricular content constraints or limits on help usage. As a result, the learner's role is best characterized as acting within a bounded decision space defined by algorithmic and contextual boundaries. The \AAFshort{} makes this distribution explicit rather than attributing outcomes solely to learner choice.

\textbf{Step 3: Describe the choice architecture.}
\revisiontwo{The third step directs researchers to describe \textit{how} the system structures the available options (e.g., defaults, constraints, salience, and transparency).} Applying the \AAFshort{} to our example, researchers describe how the tutoring system structures the help-seeking decision. The choice architecture shapes the distribution of authority: for example, \revisiontwo{surfacing a prominent hint button or making hints the default after repeated errors shifts effective control toward accepting system assistance, whereas requiring an explicit learner action to request help enhances learner decision-making.} In tutoring interfaces, help-seeking options may be surfaced as buttons, embedded in a dialogue, or triggered automatically after repeated errors. At scale, these interfaces are typically standardized, so defaults shape behavior consistently across thousands of interactions. \revisiontwo{For this focal decision, researchers would document: which options are visible (e.g., hint, retry, disengage), whether help can be refused or delayed, and how salience and defaults favor certain actions, making explicit how the design distributes authority between learner and system} \cite{brod2026agency,aleven2016help}.

\textbf{Step 4: Define evidence and accountability.}
\revisiontwo{The fourth step requires researchers to state what counts as evidence for the decision, what remains unobserved, and who is accountable for outcomes.} At-scale tutoring software generates detailed behavioral traces, including hint requests, error patterns, response timing, and conversational turns. These data allow researchers to model how learners navigate the help-seeking decision space across time. However, they do not reveal whether learners perceive help as supportive or intrusive, whether they feel able to refuse assistance, or whether reliance on hints reflects strategic choice or learned dependence. Accountability is therefore ambiguous. For example, frequent hint usage may be interpreted as gaming the system \cite{baker2006adapting}, as effective system support, or as a consequence of overly directive scaffolding \cite{fan2025beware}. \revisiontwo{The \AAFshort{} does not require a single interpretation; rather, it prompts researchers to state which interpretations are plausible, how each could be measured or tested with the available data, and how responsibility would be assigned under each, so that claims about agency are explicit and falsifiable.}

\textbf{Step 5: Set the evaluation horizon.}
\revisiontwo{The final step asks researchers to set the timescale over which consequences are evaluated (e.g., within-task, across a course), since agency-related capacities develop over time and short-term gains may mask long-term erosion.} If evaluated at the task level, automated tutoring may appear successful: error rates decrease, persistence increases, and problems are completed efficiently \cite{koedinger2015learning,koedinger2023astonishing}. When evaluated over longer horizons, however, different patterns may emerge. Repeated delegation of judgment to automated hints or LLM explanations may reduce learners' opportunities to practice monitoring their own understanding or deciding when to persist independently. At scale, these effects may be unevenly distributed, benefiting learners who already possess strong self-regulatory skills while disadvantaging others. The \AAFshort{} foregrounds the importance of aligning evaluation with the timescale over which agency-related capacities, such as help-seeking and self-monitoring, are expected to develop.

\paragraph{Implications.}
Large-scale tutoring systems do not simply support or undermine learner agency. Instead, they reconfigure where and when it is exercised. Using the \AAFshort{}, researchers can distinguish between systems that temporarily scaffold help-seeking while returning control to learners and those that permanently absorb regulatory decisions for efficiency, enabling clearer claims about whether designs foster, substitute for, or redistribute agency. In our example, help-seeking becomes a bounded interaction shaped by interfaces and automated interventions that determine when help appears, how it is framed, and how easily it can be accepted or ignored. While this often improves short-term performance—supporting persistence and task completion—it may reduce opportunities to practice judging when help is needed, navigating uncertainty, and calibrating effort. Without specifying the decision, the actors involved, and the evaluation horizon, these simultaneous effects are difficult to detect and easily misinterpreted. The case illustrates that gains in performance and engagement may reflect shifted regulatory work from learners to systems rather than enhanced self-regulation, highlighting the need to examine how decision authority is structured and redistributed over time, as centered in the \AAFshort{}.

\section{Open Challenges}

The \AAFshort{} is intended to be generative: instead of prescribing a single definition or metric of agency, it provides a structured way to surface assumptions about \emph{which} decisions matter, \emph{who} holds authority, \emph{how} choices are structured, and \emph{over what horizons} consequences unfold (Figure~\ref{fig:decision-framework-agency}). \revisiontwo{Similar to how preregistration in psychological science does not follow a single template, our aim is more to surface relevant decision points in research design for student agency, as opposed to prescribing particular sets of questions to apply the \AAFshort{}. The latter will require the development of domain- and community-specific norms.} Below, we outline several open challenges that follow from this framing and suggest synergies with adjacent communities. 

\paragraph{From Engagement to a Science of Decision Context}
A first challenge is to move from predicting engagement or performance toward a science of the \emph{decision context} under which learners can plausibly exercise agency at scale. Learning at Scale systems routinely embed automated interventions and specific forms of learner input and interaction that shape whether learners must deliberate, can meaningfully refuse assistance, or can revise goals. Yet these structures are rarely treated as primary objects of study \cite{cho2021applying}. A key research direction is to characterize which interaction forms (e.g., proactive hints, ranked recommendations, locked sequences, conversational tutoring) are agency-preserving versus agency-constraining, and to identify when such effects depend on learner characteristics, context, or institutional policy. \revisioncrc{Closely related is the challenge of operationalizing and quantifying agency from these decision-context variables in ways that are conceptually valid and produce meaningful predictions \cite{rachatasumrit2024beyond}.}

\paragraph{Leveraging Multimodal Data for Measurement}
A second challenge concerns measurement. Remote and conversational tutoring are producing increasingly rich and multimodal records (e.g., dialogue, audio, video, screen events) that can make decision points visible at fine granularity. The opportunity is to use these data to study how learners navigate choice architectures in situ: how they ask questions, resist suggestions, negotiate goals, or defer judgment to automated guidance. Multimodal data can address critiques that the field treats traces as direct evidence of intention or autonomy. \revisionthree{To move the \AAFshort{} from description to prediction or evaluation, the field needs measurable indicators (e.g., action initiation, ability to refuse or delay, decision latency), study designs that pre-specify decisions and decision-makers (e.g., cross-system comparisons), and validation strategies (e.g., inter-rater reliability in coding \AAFshort{} steps and alignment with learner self-reports of control).} Progress here requires shared measurement practices and taxonomies that explicitly separate observable decisions from unobserved constructs (e.g., endorsement, goal alignment, perceived control), and that articulate what kinds of inferences are warranted from which data sources.

\paragraph{Theory Building and Shared Taxonomies for Distributed Agency}
A third challenge is conceptual integration. Agency-related constructs are currently dispersed across literatures on SRL, help-seeking, learner control, motivation, and human–AI interaction (as was also reflected in our ad hoc literature review of past Learning at Scale research (see Section \ref{sec:lit-review})). Our field needs shared taxonomies that describe how agency is redistributed across learners, educators, institutions, and AI infrastructures in operational terms (e.g., who can choose, veto, override, contest, or audit decisions). Developing such taxonomies would enable comparability across systems and make design commitments legible, including in high-stakes contexts such as higher education where policy and regulation increasingly demand transparency and accountability \cite{pardos2026oecd}.

\paragraph{Modeling Learner Decision-Making and Agency Development Over Time}
A fourth challenge is to model agency as \emph{developmental} rather than momentary. Many systems evaluate decisions at the task level, but agency-related capacities (e.g., effective help-seeking, planning, effort regulation) emerge over weeks, months, or years \cite{bandura2006toward}. This calls for longitudinal designs and models of decision-making under constraints, including how learners adapt to system incentives and how repeated automation changes learners' propensity to deliberate, persist, or defer. Such models can connect our field to cognitive science and behavioral economics, where resource-rational decision-making and bounded cognition offer tools for describing how people act under limited time, attention, and uncertainty \cite{lieder2020resource}.

\paragraph{Integrating Agency-Preserving Design With SRL and Skill Frameworks}
Finally, a practical challenge is to translate agency claims into concrete instructional and design targets. If learners are expected to develop self-regulation skills, systems must decide which regulatory processes learners should practice and which can be safely delegated to automation. This raises design questions about creating decision points that are interpretable and consequential, and about aligning agency-preserving goals with established SRL frameworks. A productive direction might be to specify which skills (e.g., planning, monitoring, help-seeking, reflection) are being cultivated by which decision opportunities, and to evaluate whether automation supports these skills or substitutes for them.

\revisiontwo{These challenges position learner agency as a unifying research priority that connects work on learning at scale to adjacent communities in learning sciences (SRL and motivation), HCI (choice architecture and interaction design), learning analytics (multimodal measurement), cognitive science (decision modeling), and AI governance (accountability and contestability).} \revisioncrc{Multidisciplinary collaboration will be critical for validating the \AAFshort{}, by examining its utility for guiding research design and informing refinements to the framework based on emerging practice.} Taken together, the \AAFshort{} provides a shared language for asking these questions in ways that are empirically tractable at scale while remaining sensitive to the distributed, contextual nature of agency in AI-mediated learning.

\section{Limitations and Future Work}

This work has several limitations. \textit{First}, our literature review was intentionally scoped to the Learning at Scale conference proceedings and agency-related terminology, rather than providing a comprehensive review across the learning sciences, HCI, or AI ethics. Relevant adjacent work may therefore be underrepresented. Future research could extend this review, validate the \AAFshort{} against broader findings, and apply it to pathway-level decisions (e.g., course recommendation) beyond the task-level tutoring example used here.
\textit{Second}, our analysis focuses primarily on learners as focal agents in large-scale learning systems. While intentional, learner agency is embedded within broader social and organizational contexts involving teachers, tutors, families, administrators, and institutions. Future work should explicitly model how agency is distributed and negotiated across these stakeholders, particularly in K–12 and higher education settings.
\textit{In addition}, cultural and institutional norms shape how agency is expressed, valued, and constrained. Empirical studies across diverse contexts are needed to examine how agency-preserving designs function under varying expectations and how they influence persistence, self-regulation, and longer-term learning trajectories at scale.
\revisionthree{\textit{Finally}, the \AAFshort{} does not address deeper ethical questions about how power and structural inequality shape whose agency is supported or constrained. These issues extend beyond operationalization and evaluation, implicating resource allocation, institutional policy, and broader societal debate; we acknowledge them as important directions for future work.}

\section{Conclusion}

\revisionthree{As AI-mediated learning systems increasingly shape education, learner agency has become more consequential and harder to evaluate at scale. Our scoping review finds that agency remains weakly theorized, inconsistently operationalized, and often inferred from engagement or short-term performance. This gap is significant because AI systems embed assumptions about who makes decisions, which options are available, and how consequences unfold, shaping learners' capacity to act intentionally and direct their learning.}

\revisionthree{
We argue that learner agency should be treated as a first-order research concern. We identify four challenges: unclear definitions in automated settings, reliance on behavioral proxies, trade-offs among efficiency, equity, and meaningful choice, and the redistribution of agency through AI systems. To address these, we introduce the Agency Allocation Framework, which focuses on decision contexts rather than outcomes by clarifying who holds authority, which decisions matter, how choices are structured, what counts as evidence, and over what time horizons consequences are evaluated. It supports systematic study of how agency is designed, distributed, measured, and sustained in AI-mediated learning.
}

\bibliographystyle{ACM-Reference-Format}
\balance
\bibliography{main}

\end{document}